\documentclass[twocolumn,useAMS,usenatbib]{mn2e}

\usepackage{graphicx}
\usepackage{subfigure}
\usepackage{xcolor}
\usepackage{mathtext,bm,bbm,amsmath,amsfonts,amssymb,indentfirst,syntonly,graphicx}
\usepackage{mathtools}
\usepackage{slashbox}
\usepackage[english]{babel}
\usepackage{calc}
\usepackage{tikz}
\usepackage[T1]{fontenc}
\usepackage{ae,aecompl}

\def\bc{\begin{center}}
\def\ec{\end{center}}
\def\be{\begin{eqnarray}}
\def\ee{\end{eqnarray}}

\title[New constraints on the DDR from local data]{New constraints on the distance duality relation from the local data}
\author[H.-N. Lin, M.-H. Li and X. Li]
{Hai-Nan Lin$^{1}$\thanks{e-mail: linhn@cqu.edu.cn},
Ming-Hua Li$^{2}$\thanks{e-mail: liminghua@mail.sysu.edu.cn},
and Xin Li$^{1}$\thanks{e-mail: lixin1981@cqu.edu.cn}\\
$^{1}$Department of Physics, Chongqing University, Chongqing 401331, China\\
$^{2}$School of Physics and Astronomy, Sun Yat-Sen University, 135 Xingang Xi Road, Guangzhou 510275, China}
\begin{document}

\date{Accepted xxxx; Received xxxx; in original form xxxx}

\pagerange{\pageref{firstpage}--\pageref{lastpage}} \pubyear{2018}

\maketitle

\label{firstpage}

\begin{abstract}
  The cosmic distance duality relation (DDR), which connects the angular diameter distance and luminosity distance through a simple formula $D_A(z)(1+z)^2/D_L(z)\equiv1$, is an important relation in cosmology. Therefore, testing the validity of DDR is of great importance. In this paper, we test the possible violation of DDR using the available local data including type Ia supernovae (SNe Ia), galaxy clusters and baryon acoustic oscillations (BAO). We write the modified DDR as $D_A(z)(1+z)^2/D_L(z)=\eta(z)$, and consider two different parameterizations of $\eta(z)$, namely $\eta_1(z)=1+\eta_0 z$ and $\eta_2(z)=1+\eta_0 z/(1+z)$. The luminosity distance from SNe Ia are compared with the angular diameter distance from galaxy clusters and BAO at the same redshift. Two different cluster data are used here, i.e. elliptical clusters and spherical clusters. The parameter $\eta_0$ is obtained using the Markov chain Monte Carlo methods. It is found that $\eta_0$ can be strictly constrained by the elliptical clusters + BAO data, with the best-fitting values $\eta_0=-0.04\pm 0.12$ and $\eta_0=-0.05\pm 0.22$ for the first and second parametrizations, respectively. However, the spherical clusters + BAO data couldn't strictly constrain $\eta_0$ due to the large intrinsic scatter. In any case studied here, no evidence for the violation of DDR is found.
\end{abstract}


\section{Introduction}

The cosmic distance duality relation (DDR) plays an important role in cosmology and astronomy. According to this relation, the luminosity distance $D_L(z)$ is strictly correlated to the angular diameter distance by a simple formula, i.e. $D_A(z)(1+z)^2/D_L(z)=1$ \citep{Etherington:1933,Etherington:2007}. The DDR holds in any metric theory of gravity such as the general relativity, as long as the photons travel along null geodesics and the photon number is conserved during the propagation \citep{Ellis:1971,Ellis:2007}. The violation of DDR may be caused by e.g. the coupling of photon with unknown particles \citep{Bassett:2003vu}, the extinction of photon by intergalactic dust \citep{Corasaniti:2016}, the variation of fundamental constants \citep{Ellis:2013}, and so on. The modern cosmology is strongly dependent on the validity of DDR. Any violation of DDR would imply that there are new physics beyond the standard cosmological model. Therefore, testing the validity of DDR is of very importance and has aroused great interests in recent years.

A straightforward way to test the DDR is to measure the luminosity distance $D_L(z)$ and angular diameter distance $D_A(z)$ at the same redshift. A lot of works have already been done along this line \citep{Bernardis:2006,Holanda:2010vb,Piorkowska:2011nhd,Yang:2013coa,Costa:2015lja,Holanda:2016msr,Ma:2016bjt,Holanda:2016zpz,Li:2018,Hu:2018yah}. The luminosity distance is usually measured from type-Ia supernovae (SNe Ia), which are perfect standard candles and are widely used to measure the cosmological distance \citep{Perlmutter:1998np,Riess:1998mnb}. However, there is no optimal way to measure the angular diameter distance. The possible methods to determine $D_A$ include (1) by using the combined data of the $X$-ray and Sunyaev--Zeldovich (SZ) effect of galaxy clusters \citep{Filippis:2005,Bonamente:2006ct}, (2) by measuring the baryon acoustic oscillations (BAO) signal in the galaxy power spectrum \citep{Beutler:2011hx,Anderson:2013zyy,Kazin:2014qga,Delubac:2015aqe}, (3) by measuring the angular size of ultra-compact radio sources based on the approximately consistent linear size \citep{Kellermann:1993mki,Gurvits:1994fgs,Gurvits:1999hs,Jackson:2004jw}, (4) by measuring the images of quasars that are strongly gravitational lensed by foreground galaxies \citep{Cao:2015,Liao:2016uzb}, etc.

All of these methods to measure the angular diameter distance have their own advantage and shortcoming. The distance of cluster obtained using SZ effect is available in the full redshift range $z<1$, but it strongly depends on the mass profile of cluster hence induces large uncertainty. The BAO method has a high accuracy, but the number of available data points is very limited. The radio source sample is large and spans a wide redshift range, but the low redshift sources confront serious evolution effect thus are not suitable to be used as standard rulers. The strong gravitational lensing can reach to a relatively high redshift, but can only give the ratio of lens-source distance to source-observer distance, and thus signals of DDR violation, if really exists, may be partially canceled out. Until now, no evidence for the violation of DDR is found by any of these method. For example, using two different galaxy cluster data, \citet{Yang:2013coa} found that the DD relation is well compatible with observations. \citet{Ma:2016bjt} used the BAO data and found 5\% constraints in favor of DDR validity. With the ultra-compact radio source data, \citet{Li:2018} also found null result of DDR violation. \citet{Liao:2016uzb} combined the strong gravitational lensing data with cluster data and still found no evidence for the violation of DDR.

One problem of testing DDR is that the $D_L$ data and $D_A$ data are usually not measured at the same redshift. To solving this problems, several methods have been proposed, such as the nearest neighborhood method \citep{Holanda:2010vb,Liao:2016uzb}, the interpolation method \citep{Liang:2013mnf}, and the Gaussian processes \citep{Rana:2017sfr,Li:2018}. Only the data in the overlapping redshift range are available to test DDR. The furthest SNe Ia in previous datasets such as Union2.1 \citep{Suzuki:2012dhd} and JLA \citep{Betoule:2014frx} are usually bellow redshift $1.4$. Recently, a new SNe Ia sample called Pantheon is released \citep{Scolnic:2017caz}. This is the most up-to-date and largest SNe Ia sample at present. This sample consists of 1048 SNe Ia and the systematic uncertainty is reduced compared the previous compilations. Moreover, the furthest SNe reaches to redshift $z_{\rm max}\sim 2.3$ (for comparison, $z_{\rm max}\sim 1.4$ in Union2.1 and $z_{\rm max}\sim 1.3$ in JLA), which is close to the redshift of the furthest available BAO data point. So it is interesting to test DDR with the Pantheon dataset.

In this paper, we use the Pantheon compilation of SNe Ia, combined with the galaxy clusters and BAO data to test the possible violation of DDR in a model independent way. The luminosity distance of SNe Ia is first reconstructed using the Gaussian processes, then it is fitted to the combined data of clusters and BAO. The rest of the paper is organized as follows: The data samples and methodology to test DDR is illustrated in section \ref{sec:method}. The constraining results are presented in section \ref{sec:results}. Finally, discussions and conclusions are given in section \ref{sec:conclusions}.

\section{Data and Methodology}\label{sec:method}

In this section, we illustrate the method to test the DDR using the combined datasets of galaxy clusters, BAO and SNe Ia. We rewrite the possible violation of the standard DDR as
\begin{equation}\label{eq:modified_DDR}
  \frac{D_A(z)(1+z)^2}{D_L(z)}=\eta(z).
\end{equation}
Specifically, we consider two different parameterizations of $\eta(z)$, i.e.
\begin{equation}\label{eq:parametrization}
  \eta_1(z)=1+\eta_0z,~~~~\eta_2(z)=1+\eta_0\frac{z}{1+z},
\end{equation}
where $\eta_0$ is a free parameter representing the amplitude of violation of DDR. Any deviation of $\eta_0$ from zero would imply the violation of DDR. If both $D_A$ and $D_L$ have been measured at the same redshift, then the parameter $\eta_0$ can be constrained.

The first data we used to determine the angular diameter distance is the galaxy clusters. The distance of cluster can be obtained by measuring the SZ effect combined with the X-ray observation. This method to determine the angular diameter distance depends on the mass profile of clusters. \citet{Filippis:2005} have obtained the distance of 25 clusters in the redshift range [0.023,0.784] by assuming an elliptical profile, and we call this data cluster(E) hereafter. By assuming a spherical profile, \citet{Bonamente:2006ct} have obtained the distance of 38 clusters in the redshift range $[0.142,0.890]$, which we call cluster(S) hereafter.

\begin{figure}
\centering
  \includegraphics[width=0.5\textwidth]{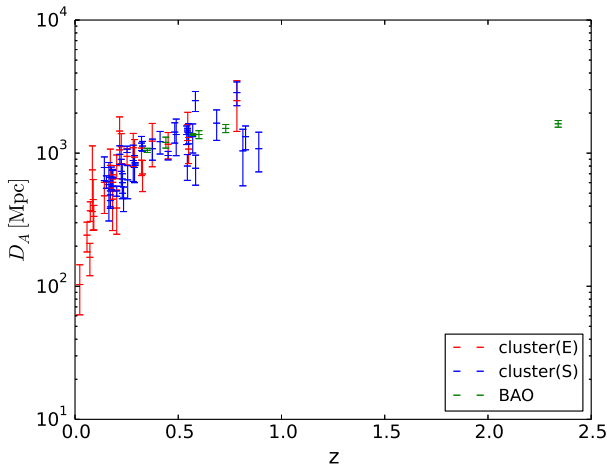}
  \caption{\small{The angular diameter distance v.s. redshift of cluster and BAO data. The elliptical cluster and spherical cluster data are presented by red and blue, respectively.}} \label{fig:DA_clus_bao}
\end{figure}

The second $D_A$ data we used come from the measurement of BAO. The BAO are regular, periodic fluctuations in the density of the visible baryonic matter, and it is widely used as the ``standard ruler" to measure the distance in cosmology. Here we use three data points from the WiggleZ Dark Energy Survey at  effective redshifts $z=0.44$, $0.6$ and $0.73$ \citep{Blake:2012mbf}, one data point from the SDSS DR7 at effective redshift $z=0.35$ \citep{Xu:2013ljd}, one data points from the SDSS-III Baryon Oscillation Spectroscopic Survey at effective redshift $z=0.57$ \citep{Samushia:2014kud}, and one data points from BOSS DR11 at effective redshift $z=2.34$ \citep{Delubac:2015aqe}. These data points are listed in Table \ref{tab:bao}, and are plotted together with the cluster(E) and cluster(S) data points in Figure \ref{fig:DA_clus_bao}.
\begin{table}
\caption{\small{The BAO data points used in this work.}}
\begin{tabular}{lll}
  \hline\hline
  $z$ & $D_A$ [Mpc] & Reference \\
  \hline
   0.44 & $1205\pm 114$ & \citet{Blake:2012mbf} \\
   0.60 & $1380\pm 95$  & \citet{Blake:2012mbf} \\
   0.73 & $1534\pm 107$ & \citet{Blake:2012mbf} \\
   0.35 & $1050\pm 38$  & \citet{Xu:2013ljd} \\
   0.57 & $1380\pm 23$  & \citet{Samushia:2014kud} \\
   2.34 & $1662\pm 96$  & \citet{Delubac:2015aqe} \\
  \hline
\end{tabular}\label{tab:bao}
\end{table}

Since the cluster and BAO data have the same format, we combined these two datasets together and calculate the luminosity distance from equation (\ref{eq:modified_DDR}),
\begin{equation}
  D_{L,{\rm clus/bao}}=\frac{(1+z)^2D_{A,{\rm clus/bao}}}{\eta(z)}.
\end{equation}
Then we convert the luminosity distance to the dimensionless distance modulus by
\begin{equation}\label{eq:mu_rs}
  \mu_{\rm clus/bao}=5\log_{10}\frac{D_{L,{\rm clus/bao}}}{\rm Mpc}+25.
\end{equation}
The uncertainty of $\mu_{\rm clus/bao}$ is obtained using the standard error propagation formula,
\begin{equation}\label{eq:mu_rs_error}
  \sigma_{\mu_{\rm clus/bao}}=\frac{5}{\ln 10}\frac{\sigma_{D_{A,{\rm clus/bao}}}}{D_{A,{\rm clus/bao}}}.
\end{equation}

The data used to determine the luminosity distance are SNe Ia. SNe Ia are widely accepted as the standard candles to measure cosmological distance thanks to the approximately consistent absolute luminosity \citep{Riess:1998mnb,Perlmutter:1998np}. The distance moduli of SNe can be calculated from the light curves using the empirical relation \citep{Tripp:1998,Guy:2005,Guy:2007}
\begin{equation}\label{eq:mu_sn}
  \mu_{\rm sn}=m_B^*+\alpha X_1-\beta \mathcal{C}-M_B.
\end{equation}
where $m_B^*$ is the B-band apparent magnitude, $X_1$ and $\mathcal{C}$ are respectively the stretch factor and color parameter, and $M_B$ is the absolute magnitude. The two nuisance parameters $\alpha$ and $\beta$ can be obtained by fitting simultaneously with cosmological parameters to a specific cosmological model. Up to now, several datasets have been released \citep{Kowalski:2008ez,Guy:2010,Amanullah:2010,Suzuki:2012dhd,Betoule:2014frx}. Recently, \citet{Scolnic:2017caz} released a new dataset called Pantheon, which consists of 1048 SNe Ia in the redshift range $0.01<z<2.3$. This is the largest SNe Ia sample released yet. The number of SNe Ia in Pantheon is about twice of that in Uinon2.1 compilation, and is about $40\%$ larger than the JLA compilation. The authors have made new efforts to reduce the systematics. Moreover, the highest redshift in Pantheon reaches to $z_{\rm max}\sim 2.3$, compared to $z_{\rm max}\sim 1.4$ in the previous datasets. Therefore, in this paper we use the new Pantheon dataset to calculate the luminosity distance.

Usually, the nuisance parameters $\alpha$ and $\beta$ are regarded as free parameters and are constrained together with cosmological parameters. However, this method strongly depends on a specific cosmological model, therefore the distance calibrated in one cosmological model couldn't be used to constrain the other cosmological models. To avoid this problem, \citet{Kessler:2017uwi} proposed a new method called BEAMS with Bias Corrections (BBC) to calibrated the SNe. The BBC method relies heavily on the method proposed by \citet{Marriner:2011} but includes extensive simulations to correct the SALT2 light curve fitter. According to the BBC method, the SNe data are binned into several redshift bins, the nuisance parameters $\alpha$ and $\beta$ are determined by fitting to a randomly chosen reference cosmology with the cosmological parameters fixed. The key point is that within each redshift bin, the local shape of the Hubble diagram is well described by the reference cosmological model. \citet{Marriner:2011} has shown that the fitted $\alpha$ and $\beta$ will converge to consistent values which are independent of the reference cosmology, as long as the bin number is large enough. Once $\alpha$ and $\beta$ are determined, a distance bias correction term $\Delta_B$ determined from simulation is added to equation (\ref{eq:mu_sn}) for each SN. The simulation also depends on an input cosmology, but the changes in the input cosmology within typical statistical uncertainties have in general a negligible effect. The Pantheon dataset is calibrated using the BBC method. \citet{Scolnic:2017caz} reported the corrected apparent magnitude $m_{B,{\rm corr}}^*=m_B^*+\alpha X_1-\beta \mathcal{C}+\Delta_B$ for all the SNe. Therefore, to calculate the distance moduli we just need to subtract $M_B$ from $m_{B,{\rm corr}}^*$ and don't need to do the color and stretch corrections any more. The statistical uncertainty $D_{\rm stat}$ and the matrix of systematics $C_{\rm sys}$ are also reported. The total uncertainty is the sum of $D_{\rm stat}$ and $C_{\rm sys}$. The detailed information on the Pantheon dataset can be found in \citet{Scolnic:2017caz}.

\begin{figure}
\centering
  \includegraphics[width=0.5\textwidth]{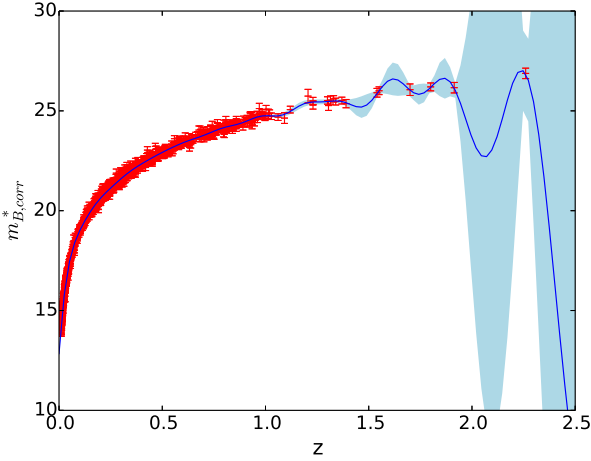}
  \caption{\small{The corrected absolute magnitudes v.s. redshift of Pantheon SNe Ia, together with the curve and $1\sigma$ uncertainty reconstructed from Gaussian processes.}} \label{fig:GP_pantheon}
\end{figure}

The SNe Ia and clusters or BAO are usually not measured at the same redshift. To solve this problem, we first use the Gaussian processes to construct the $\mu-z$ relation from the Pantheon data, then calculate the luminosity distance at the redshifts of clusters and BAO from the reconstructed $\mu-z$ relation. Unlike the best fitting method which needs an explicit fitting model, the Gaussian processes can construct a function from discrete data points without involving any model \citep{Seikel:2012uu}. The Gaussian processes only depends on the covariance function $k(x,\tilde{x})$, which characterizes the correlation between the function value at $x$ to that at $\tilde{x}$. There are many covariance functions available, but any covariance function should be positive definite and monotonously decreasing with the increment of distance between $x$ and $\tilde{x}$. Here we use the widely used squared-exponential covariance function, which reads
\begin{equation}
  k(x,\tilde{x})=\sigma_f^2\exp\left[-\frac{(x-\tilde{x})^2}{2l^2}\right].
\end{equation}
The hyperparameters $\sigma_f$ and $l$ characterize the ``\,bumpiness" of the function and should been properly chosen. We optimize the hyperparameters by maximizing the marginal likelihood marginalized over function values $f$ at the whole locations $X$. We use the publicly available python package \textsf{GaPP} \citep{Seikel:2012uu} to reconstruct the corrected apparent magnitude $m_{B,{\rm corr}}^*$ as a function of redshift. The results are plotted in Figure \ref{fig:GP_pantheon}. In the range where data points are sparse, the uncertainty of the reconstructed function is large.

The reconstructed $\mu(z)$ function is then fitted to the combined clusters and BAO data. We use the Markov chain Monte Carlo methods \citep{ForemanMackey:2012ig} to calculate the posterior probability distribution functions (pdf) of free parameters. The likelihood is given by
\begin{equation}\label{eq:likelihood}
  \mathcal{L}({\rm Data}|{\bm\theta})=\prod\frac{1}{\sqrt{2\pi}\sigma_\mu}\exp\left[-\frac{1}{2}\left(\frac{\mu_{\rm sn}-\mu_{\rm clus/bao}}{\sigma_\mu}\right)^2\right],
\end{equation}
where
\begin{equation}\label{eq:dmu_total}
  \sigma_\mu=(\sigma_{\mu_{\rm sn}}^2+\sigma_{\mu_{\rm clus/bao}}^2+\sigma_{\rm int}^2)^{1/2}
\end{equation}
is the total uncertainty inherited from SNe and cluster/BAO, ${\bm\theta}=(\eta_0, M_B, \sigma_{\rm int})$ is the set of free parameters, and the product runs over all the cluster and BAO data points. We have added an intrinsic scatter term to account for any other uncertainties. The posterior pdf is proportional to the product of likelihood and prior,
\begin{equation}
  P({\bm\theta}|{\rm Data})\propto\mathcal{L}({\rm Data}|{\bm\theta})\times P_0({\bm\theta}).
\end{equation}
We assume a non-informative prior (namely the flat prior) on all the free parameters. To ensure that equation (\ref{eq:parametrization}) is positive definite in the full redshift range of available data, we restrict $\eta_0>-0.4$. The intrinsic scatter is restricted to be $\sigma_{\rm int}>0$, and no bounds on $M_B$ are given.

\section{Results}\label{sec:results}

\begin{figure}
\centering
  \includegraphics[width=0.5\textwidth]{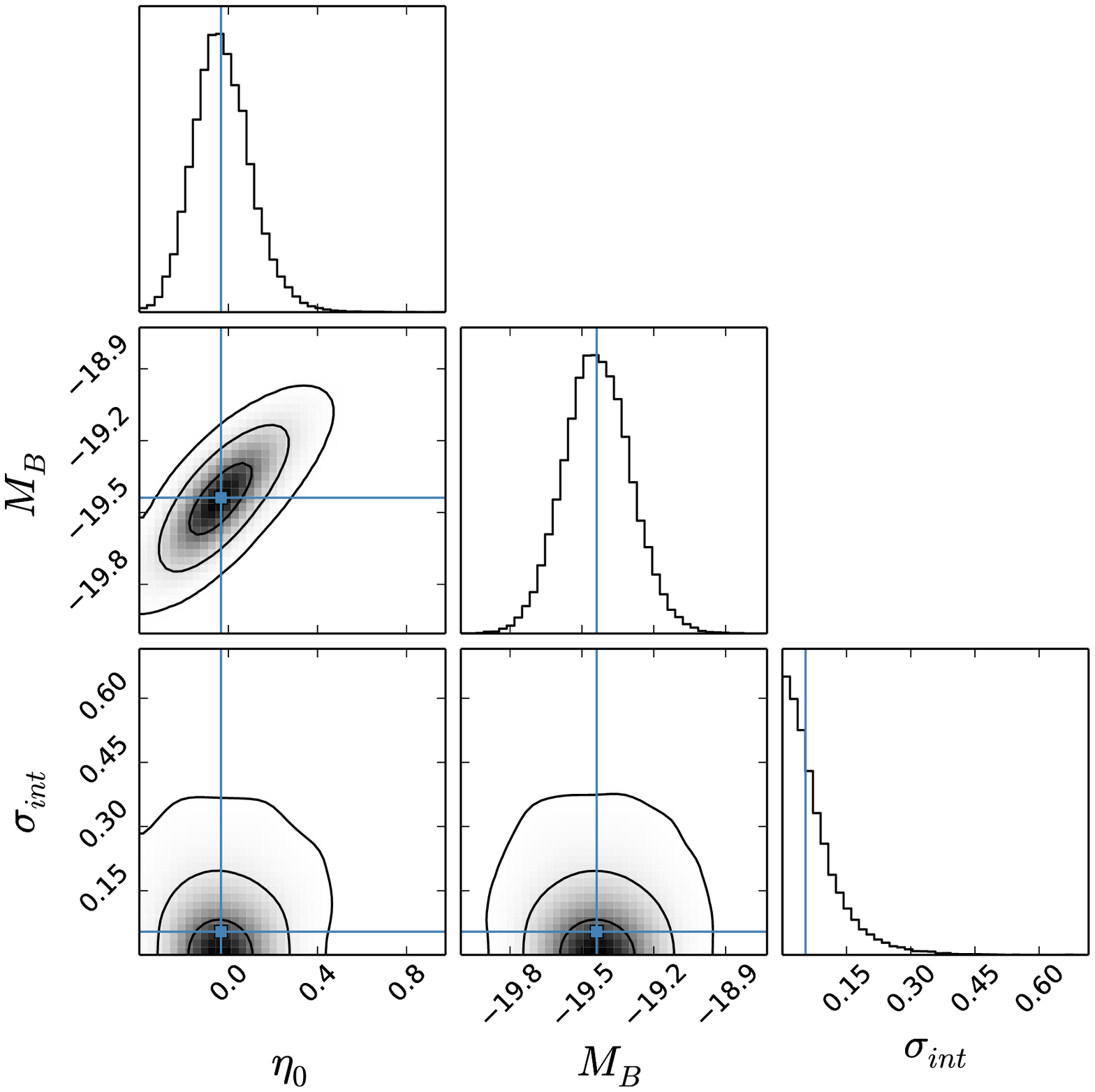}
  \caption{\small{The marginalized likelihood distributions and 2-dimensional confidence regions for the parameters $\eta_0$, $M_B$ and $\sigma_{\rm int}$ constrained from the cluster(E)+BAO data in the first parametrization $\eta_1(z)=1+\eta_0 z$.}}\label{fig:clus25_bao_p1}
\end{figure}

\begin{figure}
\centering
  \includegraphics[width=0.5\textwidth]{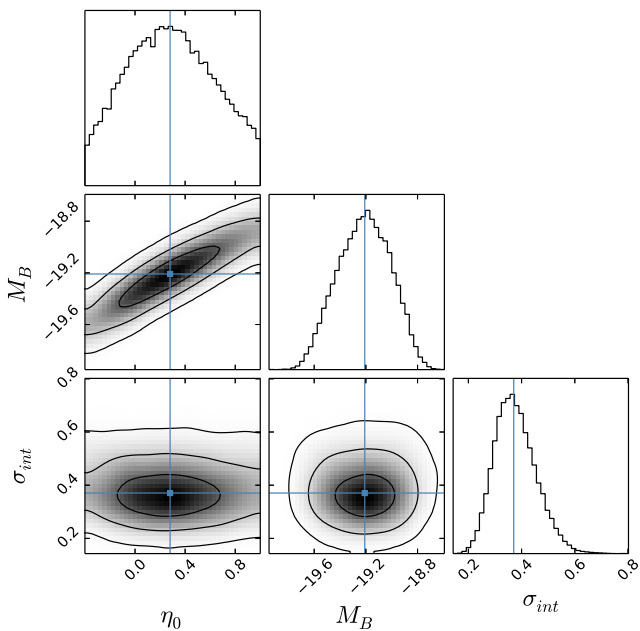}
  \caption{\small{The marginalized likelihood distributions and 2-dimensional confidence regions for the parameters $\eta_0$, $M_B$ and $\sigma_{\rm int}$ constrained from the cluster(S)+BAO data in the second parametrization $\eta_2(z)=1+\eta_0 z/(1+z)$.}}\label{fig:clus38_bao_p2}
\end{figure}

We consider two different combination of datasets, i.e., cluster(E)+BAO and cluster(S)+BAO. Each combination of dataset is used to constrain the two different parametrizations of DDR. The publicly available python package \textsf{emcee} \citep{ForemanMackey:2012ig} is used to calculate the posterior pdf of free parameters. We report the median values and the 68\% ($1\sigma$) confidence intervals of free parameters in Table \ref{tab:parameter}. The marginalized likelihood distributions and the 2-dimensional confidence regions for the parameters are plotted in Figures \ref{fig:clus25_bao_p1}\---\ref{fig:clus38_bao_p2}.

\begin{table}
\caption{\small{The best-fitting parameters and $1\sigma$ uncertainties in two different parametrizations of $\eta(z)$.}}
\begin{tabular}{ccccc}
  \hline\hline
  par. & data  & $\eta_0$ & $M_B$ & $\sigma_{\rm int}$\\
  \hline
  $\eta_1(z)$ & clus(E)+BAO & $-0.04\pm 0.12$ & $-19.44\pm 0.13$ & $<0.08$ \\
              & clus(S)+BAO & $0.13\pm 0.19$ & $-19.25\pm 0.18$ & $0.38\pm 0.08$ \\
  \hline
  $\eta_2(z)$ & clus(E)+BAO & $-0.05\pm 0.22$ & $-19.44\pm 0.16$ & $<0.08$ \\
              & clus(S)+BAO & $0.27\pm 0.36$ & $-19.21\pm 0.21$ & $0.38\pm 0.08$ \\
  \hline
\end{tabular}\label{tab:parameter}
\end{table}

\begin{figure}
\centering
  \includegraphics[width=0.5\textwidth]{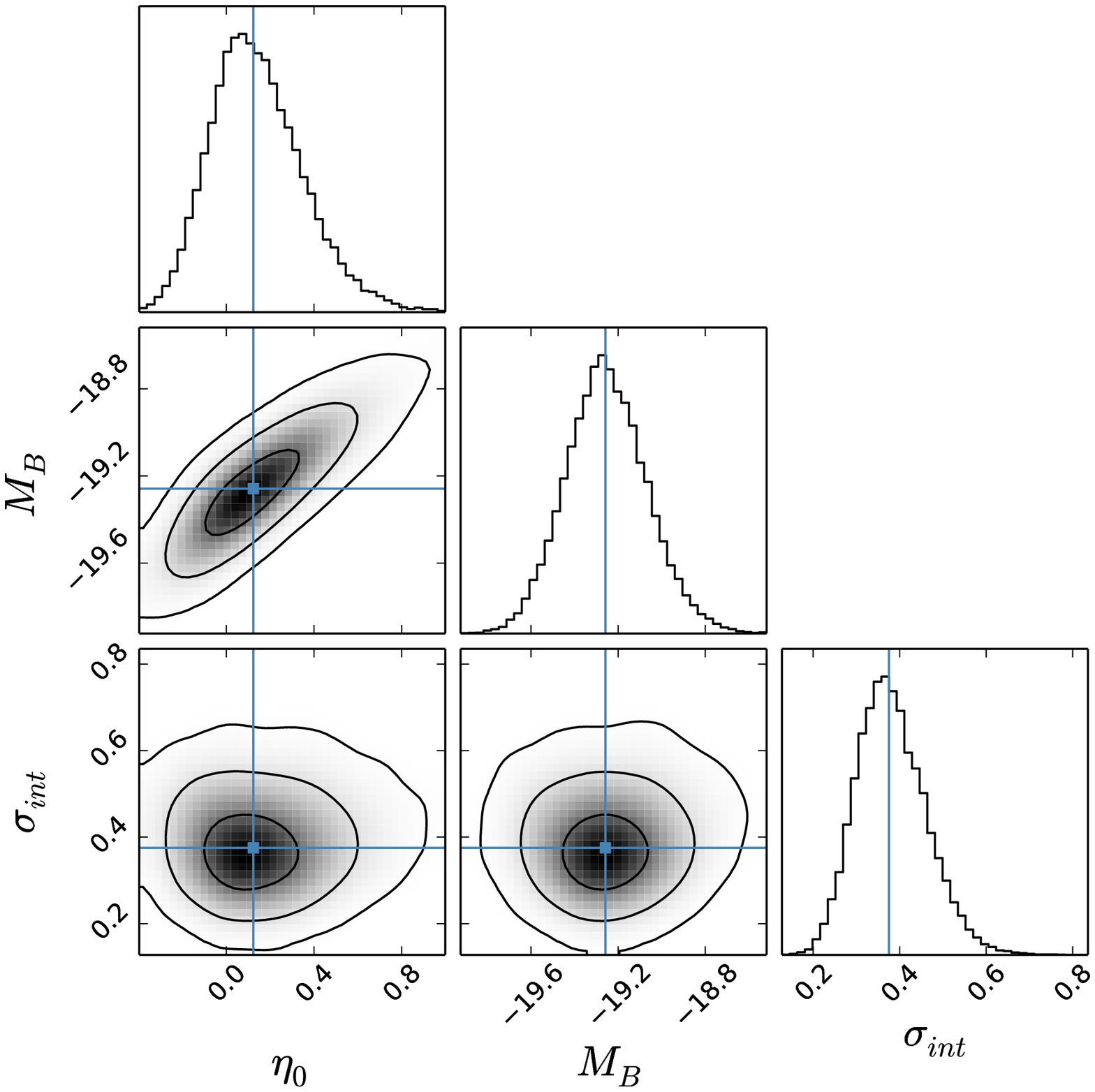}
  \caption{\small{The marginalized likelihood distributions and 2-dimensional confidence regions for the parameters $\eta_0$, $M_B$ and $\sigma_{\rm int}$ constrained from the cluster(S)+BAO data in the first parametrization $\eta_1(z)=1+\eta_0 z$.}}\label{fig:clus38_bao_p1}
\end{figure}
\begin{figure}
\centering
  \includegraphics[width=0.5\textwidth]{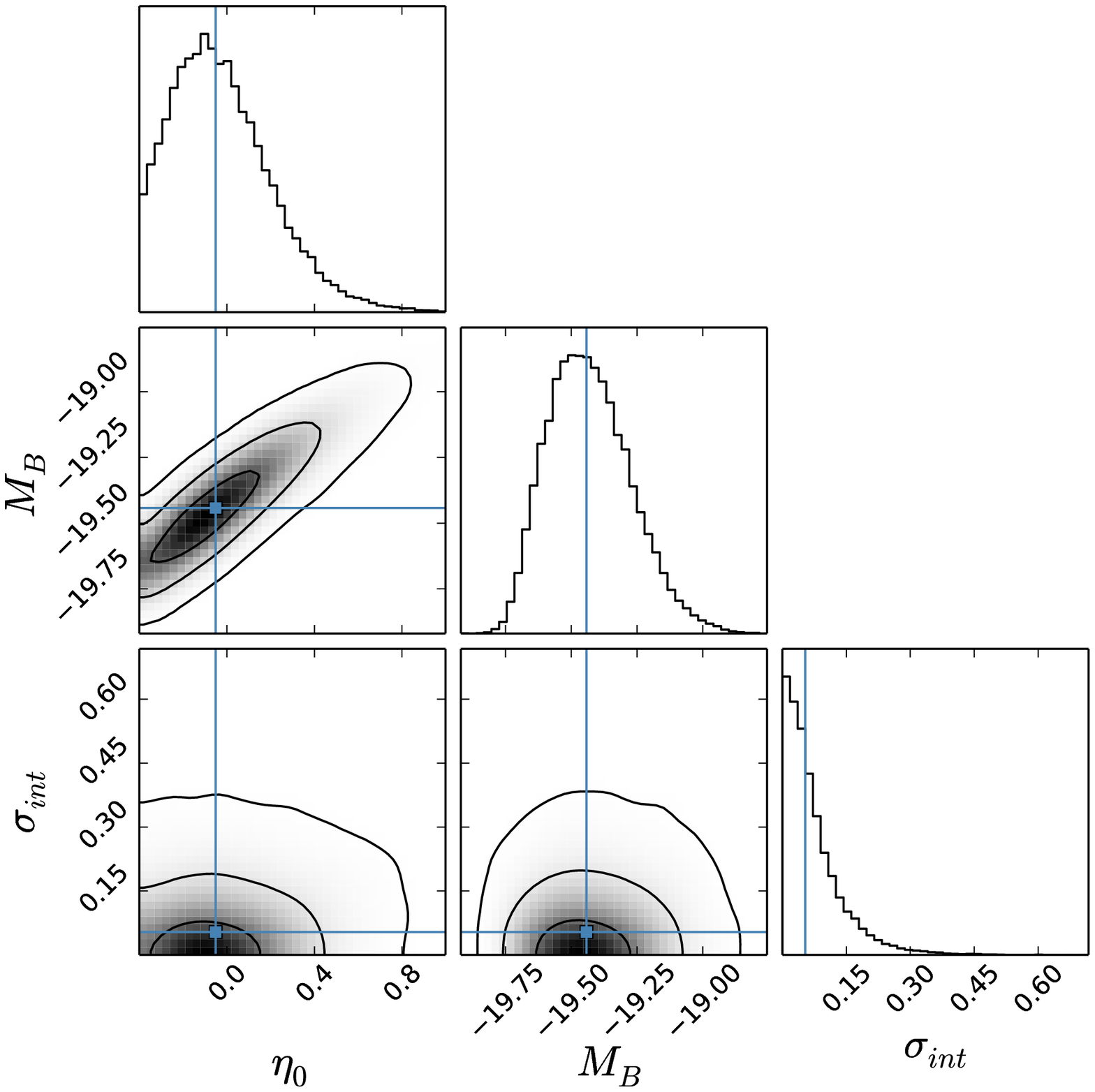}
  \caption{\small{The marginalized likelihood distributions and 2-dimensional confidence regions for the parameters $\eta_0$, $M_B$ and $\sigma_{\rm int}$ constrained from the cluster(E)+BAO data in the second parametrization $\eta_2(z)=1+\eta_0 z/(1+z)$.}}\label{fig:clus25_bao_p2}
\end{figure}

The constraints on the first parametrization of DDR from cluster(E)+BAO data is shown in Figure \ref{fig:clus25_bao_p1}. In this case, the best-fitting parameters and their $1\sigma$ uncertainties are $\eta_0=-0.04\pm 0.12$, $M_B=-19.44\pm 0.13$, $\sigma_{\rm int}<0.08$. Although $\eta_0$ and $M_B$ can be tightly constrained, only the upper limits of $\sigma_{\rm int}$ can be obtained. This means that the intrinsic scatter is negligible, and the data is well consistent with DDR.

The constraints on the first parametrization of DDR from cluster(S)+BAO data is shown in Figure \ref{fig:clus38_bao_p1}. In this case, the best-fitting parameters and their $1\sigma$ uncertainties are $\eta_0=0.13\pm 0.19$, $M_B=-19.25\pm 0.18$, $\sigma_{\rm int}=0.38\pm 0.08$. Similar to the cluster(E)+BAO data, the cluster(S)+BAO data are still well consistent with DDR. However, the intrinsic scatter is very large compared to the former.

The constraints on the second parametrization of DDR from cluster(E)+BAO data is shown in Figure \ref{fig:clus25_bao_p2}. The best-fitting parameters are consistent with the first parametrization case: $\eta_0=-0.05\pm 0.22$, $M_B=-19.44\pm 0.16$, $\sigma_{\rm int}<0.08$. The intrinsic scatter is still smaller enough to be negligible, and the DDR is well holds.

The constraints on the second parametrization of DDR from cluster(S)+BAO data is shown in Figure \ref{fig:clus38_bao_p2}. The best-fitting parameters and their $1\sigma$ uncertainties are $\eta_0=0.27\pm 0.36$, $M_B=-19.21\pm 0.21$, $\sigma_{\rm int}=0.38\pm 0.08$. The best-fitting $\eta_0$ value is a little larger than, but is still consistent with the first parametrization. As is in the first parametrization case, the intrinsic scatter of cluster(S)+BAO data is also large. Similarly, no evidence for the violation of DDR is found.

In summary, DDR can be tightly constrained by cluster(E)+BAO dataset in both parametrizations, and the intrinsic scatter is negligible. However, the cluster(S)+BAO dataset has large intrinsic scatter, and the uncertainty on the parameters is relatively large compared to the cluster(E)+BAO dataset. In any case, no evidence for the violation of DDR is found.

\section{Discussions and conclusions}\label{sec:conclusions}

In this paper, we used the SNe Ia, combined with the galaxy clusters and BAO data to constrain the possible violation of DDR. We first reconstructed the $\mu-z$ relation from the most up-to-date Pantheon compilation of SNe Ia, and then used the reconstructed $\mu-z$ relation to fit to the $D_A$ data obtained from clusters and BAO. Since the angular diameter distance measured from clusters depends on the mass profile of cluster, two different mass profiles have been used, i.e. the elliptical cluster and spherical cluster. The former contains 25 data points and the latter contains 38 data points. It was showed that the cluster(E)+BAO data have a negligible intrinsic scatter and the DDR violation can been tightly constrained in both parametrizations, i.e. $\eta_0=-0.04\pm 0.12$ and $-0.05\pm 0.22$ in the first and second parametrizations, respectively. In both parametrizations, no signal of the DDR violation was found. On the other hand, although the cluster(S)+BAO dataset is also consistent with the DDR, the intrinsic scatter of this dataset is large. This may imply that the spherical profile is not a good approximation to model the mass distribution of galaxy clusters.

We note that there is only one BAO data point at redshift $z=2.34$, while the other BAO and cluster data points all locate at redshift $z<1.0$. There is a wide redshift range between $1.0<z<2.3$ lacking of data. Therefore, it is interesting to test if the $z=2.34$ BAO data point has some influence on the results. We redo the previous calculations but omitted the $z=2.34$ BAO data point. We find that the results are almost unaffected. This is not surprising, because the reconstructed $\mu-z$ function has large uncertainty at $z=2.34$, therefore the $z=2.34$ data point has small weight in the fitting. We have also tried to fill the redshift gap between $1.0<z<2.3$ with the binned ultra-compact radio source data from \citet{Li:2018}. However, adding the radio source introduces an additional parameter (the linear size of the radio source) so couldn't help to improve the constraints.

\section*{Acknowledgements}
This work has been supported by the National Natural Science Fund of China under grant nos. 11603005 and 11775038. We are grateful to Y. Sang for useful discussions.

\label{lastpage}

\end{document}